\documentclass[12pt]{article}
\usepackage{amsmath,amssymb,amscd,cite}
\newtheorem{thm}{Theorem}[section]
\newtheorem{prop}[thm]{Proposition}
\newtheorem{lem}[thm]{Lemma}
\newtheorem{cor}[thm]{Corollary}
\newtheorem{defi}[thm]{Definition}
\newcommand{\pf}{{\bf Proof. \ }}
\newcommand{\qed}{\hfill $\Box$ \\}
\font\msbm=msbm10 at 12pt
\newcommand{\Z}{\mbox{\msbm Z}}
\newcommand{\ZZ}{\mbox{\msbm Z}}
\newcommand{\NN}{\mbox{\msbm N}}
\newcommand{\FF}{\mbox{\msbm F}}
\newcommand{\F}{\mbox{\msbm F}}
\newtheorem{rem}[thm]{Remark}
\newtheorem{ex}[thm]{Example}

\begin{document}

\title{MDS and Self-dual Codes over Rings}

\author{Kenza Guenda and T. Aaron Gulliver
\thanks{K. Guenda is with the Faculty of Mathematics USTHB, University
of Science and Technology of Algiers, Algeria. T. A. Gulliver is
with the Department of Electrical and Computer Engineering,
University of Victoria, PO Box 3055, STN CSC, Victoria, BC, Canada
V8W 3P6. email: agullive@ece.uvic.ca, tel: 250-721-6028, fax:
250-721-6052.}} \maketitle

%\normalsize

%\newpage

%\newpage
\begin{abstract}
In this paper we give the structure of constacyclic codes over
formal power series and chain rings. We also present necessary and
sufficient conditions on the existence of MDS codes over principal
ideal rings. These results allow for the construction of infinite
families of MDS self-dual codes over finite chain rings, formal
power series and principal ideal rings.
\end{abstract}

\hfill
\section{Introduction}
Although codes over rings are not new~\cite{blake}, they have
attracted significant attention from the scientific community only
since 1994, when Hammons et al. ~\cite{sole1} established a
fundamental connection between non-linear binary codes and linear
codes over $\ZZ_4$. In~\cite{sole1}, it was proven that some of the
best non-linear codes, such as the Kerdock, Preparata, and Goethal
codes can be viewed as linear codes over $\ZZ_4$ via the Gray map
from $\ZZ_4^n$ to $\FF_2^{2n}$. The link between self-dual codes and
unimodular lattices was given by Bonnecaze et al.~\cite{bonnecaze}
for $\Z_4$, and generalized by Bannai et al.~\cite{bannai}. These
results created a great deal of interest in self-dual codes over a
variety of rings, see \cite{RS} and the references there. in
Calderbank and Sloane~\cite{CS} gave the structure of cyclic codes
over $\ZZ_{p^a}$, and Kanwar, Dinh and Lopez-Permounth
\cite{K-L,permounth} presented the structure of cyclic and
negacyclic codes over chain rings. Norton and S\u al\u
agean~\cite{Ana,Norton} provided a different approach to the study
of these codes, and they considered the problem of determining the
minimum distance.

Dougherty et al.~\cite{CRT,dkk} used the Chinese remainder theorem
to generalize the structure of codes over principal ideal rings.
They gave conditions on the existence of self-dual codes over
principal ideal rings in~\cite{CRT}, and conditions on the existence
of MDS codes over these ring in ~\cite{dkk}. More recently,
Dougherty et al.~\cite{doughertyself} introduced the $\gamma-$adic
codes over a formal power series ring. The lift and projection of
these codes were also considered. In~\cite{power}, Dougherty and Liu
studied cyclic and negacyclic codes over these rings.

Recently, Dougherty~\cite{steven} posed a number of problems
concerning codes over rings. Several of these are answered in this
paper. In particular, we give necessary and sufficient conditions on
the existence of MDS codes over principal ideal rings. The existence
of such codes requires the existence of MDS codes over all the base
fields. We also give the structure of constacyclic codes over formal
power series and chain rings. The projection and the lift of these
codes is described using a generalization of the Hensel lift Lemma
and the structure of the ideals of $R[x]/\langle x^n-\lambda
\rangle$. Finally, infinite families of MDS self-dual codes are
given over principal ideal rings, finite chain rings and formal
power series.

We begin by reviewing and extending the necessary results on finite
chain rings. The lift and projection of this rings are given in the
references above. In Section 3, we give a necessary and sufficient
condition on the existence of MDS codes over principal ideal rings.
We also construct Reed-Solomon codes over these rings. In Section 4,
constacyclic codes over finite chain rings and formal power series
are examined. The structure of the ideals of $R[x]/\langle
x^n-\lambda \rangle$ is given. We consider the free constacyclic
codes and their lifts, and the number of such codes is determined.
In the last section, two families of MDS self-dual codes over chain
rings and principal ideal rings are constructed. These codes are
derived from the MDS and self-dual codes given in~\cite{guenda11}. A
table of these codes is given which includes self-dual MDS codes
derived from~\cite{bet,elias,guenda11}.

\section{Codes over Finite Chain Rings and Formal Power Series Rings}
A finite chain ring is a finite commutative ring $R$ with $1 \neq
0$, and such that its ideals are linearly ordered by inclusion.
% The ring $R$ is called a local ring if $R$ has a unique maximal ideal.
A finite commutative ring is a finite chain ring if and only if it
is a local principal ideal ring~\cite[Proposition 2.1]{permounth}.
Let $\mathfrak m$ be the maximal ideal of the finite chain ring $R$.
Since $R$ is a principal ideal ring, there exists a generator
$\gamma \in R$ of $\mathfrak m$. Then $ \gamma $ is nilpotent with
nilpotency index some integer $e$. Hence ideals of $R$ form the
following chain
$$
\langle  0 \rangle = \langle\gamma^e\rangle \subsetneq
\langle\gamma^{e-1}\rangle \subsetneq \ldots \subsetneq
\langle\gamma\rangle \subsetneq R.
$$

The nilradical of $R$ is then $\langle\gamma\rangle$, so then all
the elements of $\langle\gamma\rangle$ are nilpotent. Hence the
elements of $R\setminus \langle\gamma\rangle$ are units. Since
$\langle\gamma\rangle$ is a maximal ideal, the residue ring
$R/\langle\gamma\rangle$ is a field which we denote by $K$. This
implies that $K[X]$ is a unique factorization domain. The canonical
surjective ring morphism from $R$ to $K$ is denoted by $(-)$ and is
extended to $R[x]$ and $K[x]$.
 as follows
\begin{equation}
\label{eq:over}
\begin{split}
-:  R[X]&  \longrightarrow K[X] \\
f &\longmapsto \overline{f}=f \pmod \gamma
\end{split}
\end{equation}

Let $|R|$ denote the cardinality of $R$, and $R^*$ the
multiplicative group of all units in $R$. We know that the residue
field $K$ has characteristic $p$ and cardinality $|K|=q=p^r$ for
some integer $r$. The following Lemma is well known (see
\cite{power,doughertyself,Norton}, for example).
\begin{lem}\label{expression}
Let $R$ be a finite chain ring with maximal ideal
$\langle\gamma\rangle.$
% For any $r\in R\setminus 0$ there is a unique integer $i,\, 0\le i< e$
%such that $r=\mu\gamma^i$, with $\mu$ a unit. The unit $\mu$ is
%unique modulo $\gamma^{e-i}$.
 Let $V\subseteq R$ be a set of
representatives for the equivalence classes of $R$ under congruence
modulo $\gamma$. Then
\begin{itemize}
\item[(i)] for all $v\in R$ there exist unique $v_0,\ldots,v_{e-1}\in V$
such that $v=\sum_{i=0}^{e-1}v_i\gamma^i$;

\item[(ii)] $|V|=|K|$;

\item[(iii)] $|\langle\gamma^j\rangle|=|K|^{e-j}$ for $0\le j\le e-1$.
\end{itemize}
\end{lem}
By Lemma~\ref{expression}, we can compute the cardinality of $R$ as
follows
\begin{equation}\label{cardinality-of-R}
|R|=|K|\cdot|\langle\gamma\rangle|=|K|\cdot |K|^{e-1}=|K|^e=p^{er}.
\end{equation}

A code $\mathcal{C}$ of length $n$ over $R$ is a subset of $R$. If
the code is a submodule we say that the code is linear. Here, all
codes are assumed to be linear. If $n$ is the length of the code and
$p$ is the characteristic of $K$ we also assume that $gcd(n,p)=1$.

We attach the standard inner product to the ambient space, i.e.,
${v}\cdot {w} = \sum v_iw_i$. The dual code $\mathcal{C}^\perp$ of
$C$ is defined by
\begin{equation}
\mathcal{C}^\perp=\{ {v} \in R^n \  | \ {v} \cdot {w}= 0 {\rm \ for\
all\ } {w} \in \mathcal{C}\}.
\end{equation}
If $\mathcal{C} \subseteq \mathcal{C}^\perp$, we say that the code
is self-orthogonal, and if $\mathcal{C}=\mathcal{C}^\perp$ we say
that the code is self-dual.
% It follows from
%Proposition~\ref{prop:2.2} the following results`\cite{hand,wood}

Let $R$ be a finite chain ring. From~\cite{CS}, any linear code over
$R$ has a generator matrix in the following standard form
\begin{equation}
\label{generator} \left(  \begin{array}{ccccccc}
I_{k_0} &A_{0,1} &A_{0,2} & A_{0,3}  & \cdots& \cdots & A_{0,e-1}  \\
0       &  \gamma I_{k_1} & \gamma A_{1,2}& \gamma A_{1,3} &
\cdots& \cdots & \gamma A_{1,e-1} \\
0       & 0        & \gamma^2I_{k_2}& \gamma^2 A_{2,3} & \cdots& \cdots & \gamma^2 A_{2,e-1} \\
\vdots  & \vdots   & 0       & \ddots   & \ddots&        & \vdots     \\
\vdots  & \vdots   & \vdots  & \ddots   & \ddots& \ddots & \vdots     \\
0       & 0        & 0       & \cdots & \cdots & \gamma
^{e-1}I_{k_{e-1}} & \gamma ^{e-1}A_{e-1,e}
\end{array} \right),
\end{equation}
where the columns are grouped into blocks of sizes $ k_0,k_1,
\ldots, k_{e-1},n-\sum_{i=0}^{e-1}k_i.$ Hence $k_i$ is the number of
rows of $G$ that are divisible by $\gamma^i$, but not divisible by
$\gamma^{i+1}$. This gives that the codewords of $C$ are of the form
$(v_0,\ldots,v_{e-1})G$, where each $v_i$ is a vector of length
$k_i$ with components from $\langle\gamma^i\rangle.$
%It is well known that a finite chain ring is a Noetherian
%ring~\textbf{donner la reference de Zariski}.je me suis inspire de CS  et Norton
It follows that
\begin{equation}
\label{eq:car}
 |\mathcal{C}|=|K|^{\sum _{i=0}^{e-1}(e-i)k_i}.
 \end{equation}
 We say that $\mathcal{C}$ is of type $$1^{k_0}\gamma^{k_1}(\gamma^2)^{k_2}\ldots
 (\gamma^{e-1})^{k_{e-1}}.$$
The rank of $\mathcal{C}$ is defined to be
\begin{equation}
\label{eq:car1}
 k(\mathcal{C})=\sum_{i=0}^{e-1}k_i.
\end{equation}
It is clear that $k(\mathcal{C})$ is the minimum number of
generators of $\mathcal{C}$. Furthermore we have the following
relation between the code $\mathcal{C}$ and its dual
$\mathcal{C}^{\bot}$.
\begin{equation}
\label{wood} |\mathcal{C}||\mathcal{C}^{\bot}|=q^{\sum
(e-i)(k_i+k_i^{\bot})}=q^{en}=|R|^n, \text{ and
}(\mathcal{C}^{\bot})^{\bot}=\mathcal{C}.
\end{equation}
\begin{rem}
From (\ref{wood}), there exists a self-dual code of length $n$ over
$R$ if and only if $en$ is even. If $e$ is even, there exists a
trivial self-dual code of length $n$ given by the generator matrix
$G=\gamma ^{\frac{e}{2}}I_n.$
\end{rem}
The free rank of $\mathcal{C}$ is defined to be the maximum of the
ranks of the free submodules of $\mathcal{C}$. A linear code is said
to be free if its free rank is equal to its rank. In this case, the
code is a free $R$-submodule which is isomorphic as a module to
$R^{k(\mathcal{C})}$, and has a basis of $k(\mathcal{C})$ elements.
%Hence in this case a generator matrix of $\mathcal{C}$ in standard
%form is given by $(I_k\; M)$ for some matrix $M$.
%\begin{lem}(\cite{CS,Ana})
%\label{thm:csn} Let $C$ be a linear code with generator matrix $G$
%in standard form~(\ref{generator}). Then if for $0\leq i <j \leq e,$
%$B_{i,j}=-\sum_{k=i+1}^{j-1}
%B_{i,k}A_{e-j,e-k}^{tr}-A^{tr}_{e-j,e-i}$, we have that
%\begin{equation}H=
%\label{generator2} \left(  \begin{array}{ccccc}
% B_{0,e}& B_{0,e-1} &\dots & B_{0,1}& I_{n-k}  \\
%   \gamma B_{1,e}  & \gamma  B_{1,e-1}& \dots& \gamma I_{k_{e-1}} & 0 \\
%   \gamma^2 B_{2,e}  & \gamma  B_{2,e-1}& \dots& \gamma I_{k_{e-2}} & 0 \\
%\vdots \\
%\gamma^{e-1}B_{e-1,e}& \gamma^{e-1}I_{k_{e-1}}&\dots&0&0
%\end{array} \right),
%\end{equation}
%is a generator matrix for $C^{\bot}$. Furthermore
%$k_i(C^{\bot})=k_{e-i}(C)$, for $i=1,\ldots,e-1,$ and
%$k_0(C^{\bot})=n-k(C)$.
%\end{lem}
%From Lemmas~\ref{expression} and~\ref{thm:csn}, we obtain that
The Hamming weight of a codeword $v$ of $\mathcal{C}$ is the number
of non-zero coordinates, and for a code $\mathcal{C}$  we denote by
$d_H(\mathcal{C})$ or simply $d$ the non-zero minimum Hamming
distance of $\mathcal{C}$.
%A linear code $C$ can also be described in term of its parity check matrix
%$H$, i.e., $v\in C$ if and only if $Hc^t=0$.
%Then the minimum Hamming weight of $C$ is $d$ if and only if any $d-1$ column of $C$
%are linearly independent but some $d$ columns are not.

The well known Singleton bound for codes over any alphabet of size
$m$ (see\cite{MS}) gives that
\begin{equation}
\label{single} d_H(\mathcal{C}) \leq n- \log_m(|\mathcal{C}|) +1.
\end{equation}
If a code meets this bound, it is called maximum distance separable
(MDS). For codes over principal ideal rings we have the following
bound~\cite{shiro}
\begin{equation}
\label{MDRbound}
d_H(\mathcal{C})\le n-k(\mathcal{C})+1.
\end{equation}
This is a stronger bound in general unless the linear code is free,
in which case the bounds coincide.
 If a code over $R$ meets the bound (\ref{MDRbound}), then we say that $\mathcal{C}$ is a Maximum Distance with respect to Rank
(MDR) code.
%For a full description of these codes, see \cite{MDR}.
%\begin{prop}
%\label{wood} Let $R$ be a finite chain ring of size $p^{\alpha}$.
%The number of codewords in any linear code $C$  of length $n$ over
%$R$ is $p^k$, for some integer $0\leq k\leq \alpha n$. Moreover the
%dual code $C^{\bot}$ has $p^{\alpha n -k}$, so that
%$|C||C^{\bot}|=|R|^n.$
%\end{prop}
%\begin{thm}
%\label{thm:main0} If there exists an MDS code over $R$ of rank $k$
%and length $n$ then there exists an MDS code over $K$ of dimension
%$k$ and length $n$.
%\end{thm}
%\pf Let $C$ be an MDS code over $R$  and let $C'=\gamma^{e-1}C$,
%then the minimum Hamming weight of $C'$ is less or equal to the
%minimum Hamming weight of $C$. From Lemma~\ref{lem:necMDS} the code
%$C$ is free. Hence by~\cite[Porposition 3.11]{Norton}
%$\gamma^{e-1}C=C\cap \gamma^{e-1}R $. Hence the minimum Hamming
%distance of $C$ is less or equal to the minimum Hamming distance of
%$C'$. Let now the map $\phi : \alpha R^n \rightarrow K^n$ given by
%$\phi(\alpha c)= \overline c$, hence from~\cite[Lemma 2.4]{Norton}
%this map is an isomorphism of $K$ vector spaces. Hence the image of
%$C'$ is a $K$-linear code $B$ isomorph to $C'$. Hence $|C|'=|B|$ .
%But $|C'|=p^k$. Since $C$ is MDS then $d_H(C)=n-k+1=d_H(C')=d_B$.
%Then $B$ is an $[n,k,d_H(B)]$ linear code over $K$ which verifies
%$d_H(B)=n-k+1$. Hence we obtain that the code $B$ is an MDS code
%over $K$. \qed
The submodule quotient of $C$ by $v\in R$ is the code
$$(\mathcal{C}:v)=\{x \in R^n | xv\in \mathcal{C}\}.$$
Thus we have the tower of linear codes over $R$
\begin{equation}
\label{eq:inclusion} \mathcal{C}=(\mathcal{C}:\gamma) \subseteq
\ldots (\mathcal{C}:\gamma^i)\subseteq \ldots \subseteq
(C:\gamma^{e-1}).
\end{equation}
%Let $C$ be a linear code over $R$. Consider the codes
%$(\mathcal{C}:\gamma^i)=\{ v \ | \ \gamma^i v \in \mathcal{C} \}$,
%\, $i=0,1,\dots, e-1$, over $R$, and such that the residue fields
%are $R/\langle \gamma \rangle=K$.
%Let ``$-$'' denote the canonical map $R^n\to K^n,
%\,\,(n\geq 1)$.
For $i=1,2,\dots, e-1$ the projection of $(\mathcal{C}:\gamma^i)$
over the field $K$ are denoted by $Tor_i(\mathcal{C})
=\overline{(\mathcal{C}:\gamma^i)}$ , and called the \textit{torsion
codes} associated with the code $\mathcal{C}$. By a similar prove as
~\cite[Theorem 5.1]{optimal} one can obtain the following result.
\begin{equation}
 \label{eq:tor1} |Tor_i(C)| = \prod_{j=0}^i q^{k_j},
\end{equation}
Using (\ref{eq:inclusion}) we can obtain easily the following tower.
\begin{equation}
\label{eq:toow}
 Tor_0(C)\subset Tor_1(C)\subset \ldots \subset Tor_{e-1}(C)\subset
Tor_0(C)^{\bot}
\end{equation}
\begin{prop}
\label{th:5.4} Let $R$ be a finite chain ring with maximal ideal
$\gamma$ and nilpotency index $e$. Then the following holds:
\begin{itemize}
\item[(i)] If $\mathcal{C}$ is a linear MDS code over $R$ of rank $k=k(C)$ and type
$1^{k_0}\gamma^{k_1}(\gamma^2)^{k_2}\ldots
(\gamma^{e-1})^{k_{e-1}}$, we have that $k_i=0$ for $i>0$.
Furthermore we have $Tor_i(C)=Tor_0(C)$ for all $0 \leq i,j \leq
e-1$, and it is an MDS code of length $n$ and dimension $k$ over the
field $K$.
\item[(ii)] If there exists an MDR code over $R$, then $Tor_{e-1}(C)$ is
an MDS code over the field $K$.
%\item[(iii)] if $C$ is a
%self-orthogonal code over $\ZZ_4$, then the code $Tor_0(C)$ is a
%doubly-even self-orthogonal code,
\item[(iii)]
if $rank(C)=n/2$, then $C$ is free, $rank(C)=rank(Tor_i(C))$,
$Tor_i(C)=Tor_j(C)$, $Tor_i(C)$ is self-dual for all $0 \leq i,j
\leq e-1$.
% and
%\begin{equation}
%\label{eq:anna}
% d_{H}(Tor_i(C))=d_H(C).
%\end{equation}
\end{itemize}
\end{prop}
\pf From (\ref{eq:car}) we have $|\mathcal{C}|<p^{erk}$. If $k_i>0$
for any $i>0$, the code meets the bound given in~(\ref{single}),
which prevents the code from meeting the bound given
in~(\ref{MDRbound}). Which mean that $\mathcal{C}$ is a free code.
%Lemma~\ref{lem:anna} $Tor_i(C)=Tor_0(C)$ for all $0 \leq i,j \leq $.
From~\cite[Theorem 5.3]{dkk} $Tor_i(\mathcal{C})=Tor_j(\mathcal{C})$
for all $0 \le i,j \le e-1$ and $Tor_i(\mathcal{C})$ are MDS. Part
(ii) follows from~\cite[Theorem 5.4]{dkk}.
% To prove (iii), assume  that
%$\mathcal{C}$ is a self-orthogonal code over $\ZZ_4$. From
%\cite[Theorem~12.1.5]{huffman03}, the Euclidean weight of any
%codeword $c \in C$ satisfies
%\begin{equation}
%\label{eq:euc} wt_E(c)=n_1(c)+4 n_2(c)+n_3(c) \equiv 0 \pmod 4.
%\end{equation} Let $c^1$ be a codeword in $Tor_0(C)$. Then for $c^1
%\equiv c \pmod 2$ we have $wt_H(c^1)=n_1(c)+n_3(c)$, and from
%(\ref{eq:euc}) we obtain $wt_H(c^1) \equiv 0 \pmod 4$.
 Assume now
that $\mathcal{C}$ is self-orthogonal such that
$rank(\mathcal{C})=n/2$. Then from $(ii)$, $Tor_i(\mathcal{C})$ is
self-orthogonal for all $0\leq i \leq \lfloor \frac {e-1}{2}
\rfloor$, and $rank(Tor_0(\mathcal{C}))=n/2$. Thus
$Tor_0(\mathcal{C})$ is self-dual, and from~(\ref{eq:toow}) we have
$Tor_i(\mathcal{C})= Tor_j(\mathcal{C})$ so that $\mathcal{C}$ is
free and $rank(\mathcal{C})=rank(Tor_i(\mathcal{C}))$. \qed

%The results of Theorem~\ref{th:5.4} give that if there are MDS and
%MDR codes over a finite chain ring $R$, then there must be MDS codes
%of the same length and rank over $K$. It is well known that there
%are no non-trivial binary MDS codes. Hence if $K$ is isomorphic to
%$\F_2$, there are no non-trivial MDS or MDR codes over $R$. For
%example, there are no non-trivial MDS or MDR codes over any finite
%chain rings with residual field $\F_{2}$.
%\begin{thm}
%Let $C$ be a self-orthogonal code over $R$. Then $Tor_i(C)$
%satisfies the following
%\begin{enumerate}
%\item[(i)],
%\item[(ii)]
%$ Tor_i(C)$, is self-orthogonal for all $0\leq i \leq \left\lfloor
%\frac {e-1}{2} \right\rfloor $,

%\end{enumerate}
%\end{thm}
%\pf (i) and (ii) follow from the structure of the generator matrices
%and Part (ii) of Lemma~\ref{lem:anna}.
%
%\subsection{Lifts and Projections of Codes over Chain Rings}
Let $R$ be a finite chain ring with maximal ideal $\langle \gamma
\rangle $, nilpotency index $e$ and residue field $K$. Hence from
Lemma~\ref{expression} any element $a$ of $R$ can be written
uniquely as $a = a_0 + a_1\gamma + \dots + a_{e-1} \gamma^{e-1}$,
where $a_i \in K$. For an arbitrary positive integer $i$, we define
$R_i$ as
\begin{equation}
R_i=\{a_0+a_1\gamma + \ldots+ a_{i-1} \gamma^{i-1} | a_i \in K \},
\end{equation}
%where $\gamma^{i-1} \neq 0,$ but $\gamma^i = 0$ in $R_i$. The
%following two operations can be defined over $R_i$
%\begin{equation}
%\sum_{l=0}^{i-1}a_l\gamma^l+\sum_{l=0}^{i-1}b_l\gamma^l=\sum_{l=0}^{i-1}(a_l+b_l)\gamma^l
%\end{equation}
%\begin{equation}
%\sum_{l=0}^{i-1}a_l\gamma^l.\sum_{l'=0}^{i-1}a_{l'}\gamma^{l'}=\sum_{s=0}^{i-1}(\sum_{l+l'=s}a_lb_{l'})\gamma^s
%\end{equation}
%With the operations defined above.
 Then $R_i$ are finite chain rings with $R_1=K$ and $R_e=R$. Each $R_i$ is with index of nilpotency
$i$ and with maximal ideal $\langle\gamma
 \rangle$ and with set of unit
 \begin{equation}
\label{lem:any}
 R_i^*=\{\sum_{l=0}^{i-1} a_l\gamma^l |\, 0\neq a_0 \in K\}.
\end{equation}
The ring of formal power series $R_\infty$ is defined as follows
\begin{equation}
\label{eq:infty}
R_\infty=K[[\gamma]]=\{\sum\limits_{l=0}^{\infty}a_l\gamma^l\,|\,
a_l \in K\}.
\end{equation}
%It is well known that for any two elements
%$a=\sum\limits_{l=0}^{\infty}a_l\gamma^l$ and
%$b=\sum\limits_{l=0}^{\infty}b_l\gamma^l \in R_\infty$, their sum
%and product are the following~\cite{Bourbaki_CA,Z-S}
%\begin{eqnarray*}
%\sum\limits_{l=0}^{\infty}a_l\gamma^l+\sum\limits_{l=0}^{\infty}b_l\gamma^l
%&=&\sum\limits_{l=0}^{\infty}(a_l+b_l)\gamma^l\\
%\sum\limits_{l=0}^{\infty}a_l\gamma^l\cdot\sum\limits_{l'=0}^{\infty}b_{l'}\gamma^{l'}
%&=&\sum\limits_{s=0}^\infty(\sum\limits_{l+l'=s}a_lb_{l'})\gamma^s.
%\end{eqnarray*}
The following result is well known~\cite{Bourbaki_CA,power,Z-S}.
\begin{lem}
\label{lem:bou} Assume the notation given above. Then we have that
\begin{itemize}
\item[(i)] $R^{\times}_\infty=\{\sum\limits_{l=0}^{\infty}a_l\gamma^l\,|\,
a_0\ne 0\}$;
\item[(ii)] the ring $R_\infty$ is a principal ideal domain with a unique
maximal ideal $\langle \gamma \rangle$.
\end{itemize}
\end{lem}
\bigskip
Hence from Lemma~\ref{lem:bou}, any nonzero element $a$ of
$R_\infty$ can be written as
\begin{equation}
\label{eq:com}
 a=\gamma ^ld \text{ with }d \text{ a unit in }R_\infty.
\end{equation}

The generator matrix of a linear code over $R_\infty$ is given by
the following Lemma.
\begin{lem}(\cite[Lemma 3.3]{doughertyself}
Let $\mathcal{C}$ be a nonzero linear code over $R_\infty$ of length
$n$. Then any generator matrix of $\mathcal{C}$ is permutation
equivalent to a matrix of the following form
\begin{equation}\label{generator-1}
G=\left( \begin{array}{ccccccc}
\gamma^{m_0}I_{k_0} & \gamma^{m_0}A_{0,1} & \gamma^{m_0}A_{0,2} & \gamma^{m_0}A_{0,3} & & & \gamma^{m_0}A_{0,r} \\
       & \gamma^{m_1} I_{k_1} & \gamma^{m_1} A_{1,2}&
\gamma^{m_1} A_{1,3} & & &
\gamma^{m_1} A_{1,r} \\
       & & \gamma^{m_2} I_{k_2}& \gamma^{m_2} A_{2,3} & & & \gamma^{m_2} A_{2,r} \\
 & & & \ddots & \ddots& & \\
  & & & & \ddots& \ddots & \\
       & & & & & \gamma^{m_{r-1}}
I_{k_{r-1}} & \gamma^{m_{r-1}} A_{r-1,r}
\end{array} \right),
\end{equation}
where $0\le m_0<m_1<\cdots<m_{r-1}$ for some integer $r$.
\end{lem}
\begin{rem}
A code $\mathcal{C}$ with generator matrix of the form given in
(\ref{generator-1}) is said to be of type
$$
(\gamma^{m_0})^{k_0}(\gamma^{m_1})^{k_1}\cdots(\gamma^{m_{r-1}})^{k_{r-1}},
$$
where $k=k_0+k_1+\cdots+k_{r-1}$ is called the rank, and is the rank
of $\mathcal{C}$ as a module.
\end{rem}
A code $\mathcal{C}$ of length $n$ with rank $k$ over $R_\infty$ is
called a {\it $\gamma$-adic $[n,k]$ code}.
% We call $k$ the dimension
%of $\mathcal{C}$ and is denoted by $\dim \mathcal{C}=k$.

Since $R_\infty$ is a principal ideal ring. Hence the codes over
$R_\infty$ satisfy the bound~(\ref{MDRbound}).
%Codes meeting
%thebound~(\ref{MDRbound}) are called MDR codes over $R_\infty$.
 An MDR code over $R_\infty$ is said to be MDS if it is of type $1^k$
for some $k$. We have the following result.
\begin{thm}
\label{type}(\cite{doughertyself}) If $\mathcal{C}$ is a linear code
over $R_\infty$ then $\mathcal{C}^\bot$ has type $1^m$ for some $m$.
Furthermore, the following holds:
\begin{itemize}
\item[(i)] $\mathcal{C}=(\mathcal{C}^\bot)^\bot$ if and only if $\mathcal{C}$ has type $1^k$.
\item[(ii)] If $\mathcal{C}$ is an MDR or MDS code then $\mathcal{C}^{\bot}$ is an MDS code.
\end{itemize}
\end{thm}
%\pf Assume $\mathcal{C}$ is a code of length $n$ and rank $k$ with
%$d_H(\mathcal{C})=n-k+1$. Then we know that $\mathcal{C}^{\bot}$ is of type $1^{n-k}$.
%Since $R_\infty$ is a domain, we get that any $n-k$
%columns of the generator matrix of $\mathcal{C}^{\bot}$ are linearly
%independent. This gives that the minimum Hamming weight of
%$\mathcal{C}^{\bot}$ is $n-(n-k)+1=k+1$.\qed

%%%%%%%%%%%%%%%%%%%%%%%%%%%%%%%%%%%%%%%%%%%%%%%%%%%%%%%%%%%%%%%%%%%%%%%%%%%%%%%%%%%%%%%%%%%%%%%%%%%%%%%%%%
For two positive integers $i<j$, we define a map as follows
\begin{eqnarray}
\label{projection}\Psi^j_i:R_j &\to&R_i,\\
\sum\limits_{l=0}^{j-1}a_l\gamma^l&\mapsto&\sum\limits_{l=0}^{i-1}a_l\gamma^l.
\end{eqnarray}
If we replace $R_j$ with $R_\infty$ then we denote $\Psi^{\infty}_i$
by $\Psi_i$. Let $a$ and $b$ be two arbitrary elements in $R_j$. It
is easy to show that
\begin{equation}\label{eqau}
\Psi^j_i(a+b)=\Psi^j_i(a)+\Psi^j_i(b),\,\,
\Psi^j_i(ab)=\Psi^j_i(a)\Psi^j_i(b).
\end{equation}
If $a,b\in R_\infty$, we have that
\begin{equation}\label{eqau1}
\Psi_i(a+b)=\Psi_i(a)+\Psi_i(b),\,\, \Psi_i(ab)=\Psi_i(a)\Psi_i(b).
\end{equation}
Note that the two maps $\Psi_i$ and $\Psi_i^j$ can be extended
naturally from $R_\infty^n$ to $R_i^n$ and $R_j^n$ to $R_i^n$,
respectively.
%
%\begin{rem} The construction method above gives the following series of chain
%rings
%\[
%R_\infty \rightarrow  \ldots \rightarrow  R_e \rightarrow R_{e-1}
%\rightarrow \ldots \rightarrow  R_1 = K.
%\]
%\end{rem}
\begin{defi}
 An $[n,k]$ code $\tilde{\mathcal{C}}$ over $R_{j}$ is said to be the lift of a code $C$ over
 $R_{i}$, with $i$ and $j$ integers such that $1\le i\le j<\infty$, if
$\tilde{\mathcal{C}}$ has a generator matrix $\tilde{G}$ such that
$\Psi_i^j(\tilde{G})$ is a generator matrix of $\mathcal{C}$. Hence
we have $\mathcal{C}=\Psi_i^j(\tilde{\mathcal{C}})$. If
$\mathcal{C}$ is an $[n,k]$ $\gamma$-adic code, then for any
$i<\infty$, we call $\Psi_i(\mathcal{C})$ a projection of
$\mathcal{C}$. We denote $\Psi_i(\mathcal{C})$ by $\mathcal{C}^i$.
\end{defi}
\begin{rem} The map $\Psi_1^i$ is the same map as that given in~(\ref{eq:over}).
Hence when $\mathcal{C}$ is a cyclic code over $K$ generated by a
polynomial $g$, then the code over $R_{i}$ generated by the lifted
polynomial of $g$ is the lifted code $\tilde{\mathcal{C}}$ in the
sense of the definition above.
\end{rem}
\begin{lem}
\label{lem:frat} Let $\mathcal{C}$ be a free code over $R_i$. Then
the lifted code $\tilde{\mathcal{C}}$ of $\mathcal{C}$ over $R_j$,
$j \ge i$, is a free code.
\end{lem}
\pf If $\mathcal{C}$ is a free code of rank $k(\mathcal{C})$ over
$R_i$, then $\mathcal{C}$ is isomorphic as a module to
$R_i^{k(\mathcal{C})}$. Hence the $k$ rows of the generators matrix
$G$ of $\mathcal{C}$ are linearly independent. Since the map
$\Psi_i^j$ is a morphism, the rows of $\tilde{G}$ are also linearly
independent, otherwise the rows of $G=\Psi_i^j(\tilde{G})$ are not
linearly independent, which is absurd. It then follows that the code
$\tilde{\mathcal{C}}$ is also a free code over $R_j$. \qed

\begin{lem}
\label{th:min}(\cite[Theorem 2.11]{doughertyself}) Let $\mathcal{C}$
be a $\gamma$-adic code. Then the following two results hold.
\begin{itemize}
\item[(i)] the minimum Hamming distance $d_H(\mathcal{C}^i)$ of
$\mathcal{C}^i$ is equal to $d=d_H(\mathcal{C}^1)$ for all
$i<\infty$;
\item[(ii)] the minimum Hamming distance $d_\infty=d_H(\mathcal{C})$ of
$\mathcal{C}$ is at least $d=d_H(\mathcal{C}^1)$.
\end{itemize}
\end{lem}
\begin{thm}
\label{thliftMDS} Let $\mathcal{C}$ be a linear code over $R_i$, and
$\tilde{\mathcal{C}}$ be a lift code of $\mathcal{C}$ over $R_j$,
where $\infty \geq j>i$. If $\mathcal{C}$ is an MDS code over $R_i$
then the code $\tilde{\mathcal{C}}$ is an MDS code over $R_j$ with
the same minimum distance $d_H$.
\end{thm}
\pf Let $\mathcal{C}$ be an MDS linear code of length $n$ and
dimension $k$, so that $d_H=n-k+1$. Let $v$ be a codeword of $C$ of
minimum Hamming weight. We have that $\tilde{\mathcal{C}}$ is a
linear code over $R_i$ with length $n$ and rank $k$. The vector $v$
can be viewed as a codeword of $\tilde{\mathcal{C}}$ since we can
write $v=(v_1,\cdots,v_n)$ where
$$
v_l=a^l_0+a^l_1\gamma+\cdots+a^l_{i-1}\gamma^{i-1}+0\gamma^i+\cdots+0\gamma^{j-1}
+ \ldots .
$$
Let $w$ be any lifted codeword of $v$. Then we have that $w_H(w)\ge
w_H(v)$. On the other hand, for any lifted codeword $w'$ of $v'$,
where $ v'\in C$, we also have that $w_H(w')\ge w_H(v') \ge w_H(v)$.
Hence by~Lemma~\ref{th:min} we obtain that the minimum Hamming
weight of $\tilde{\mathcal{C}}$ is $d_H$, and this implies that
$\tilde{\mathcal{C}}$ is an MDR code for all $j>i$. From
Proposition~\ref{th:5.4} we have that an MDS code is a free code.
Hence $\mathcal{C}$ is a free code, and by Lemma~\ref{lem:frat} the
lifted code $\tilde{\mathcal{C}}$ is also free. Thus
$\tilde{\mathcal{C}}$ is an MDS code.\qed
%
%\subsection{Self-Dual $\gamma-$adic Codes}
%Let $\FF_q$ be a finite fields with characteristic $p$. The
%following theorem is well known~\cite{RS}.
%\begin{thm}\label{aa}
%\begin{itemize}
%\item[(i)] If $p=2$ or $p\equiv 1\, (\,{\rm mod}\, 4)$, then a self-dual
%code of length $n$ exists over $\FF_q$ if and only if $n\equiv 0\,
%(\,{\rm mod}\, 2)$;
%\item[(ii)] If $p\equiv 3\, (\,{\rm mod}\, 4)$, then a self-dual code of
%length $n$ exists over $\FF_q$ if and only if $n\equiv 0\, (\,{\rm
%mod}\, 4)$.
%\end{itemize}
%\end{thm}
%\bigskip
%Theorem~\ref{aa} can be extended to $R_\infty$, and then to codes
%over rings using the following theorem.
%\begin{thm}\label{self-dual-1}(\cite[Theorem  3.4]{doughertyself}
%If $\mathcal{C}$ is a self-dual code of length $n$ over $R_\infty$,
%then $\Psi_i(\mathcal{C})$ is a self-dual code of length $n$ over
%$R_i$ for all $i<\infty$.
%\end{thm}
%From Theorems~\ref{aa} and \ref{self-dual-1}, we obtain the
%following result.
%\begin{cor}(\cite[Corollary 4.5]{doughertyself}
%Let $\mathcal{C}$ be a self-dual code of length $n$ over $R_\infty$
%with residue field $K$. If $p$ is the characteristic of $K$, then we
%have
%\begin{itemize}
%\item[(i)] If $p=2$ or $p\equiv 1\, (\,{\rm mod}\, 4)$, then $n\equiv 0\,
%(\,{\rm mod}\, 2)$;
%\item[(ii)] If $p\equiv 3\, (\,{\rm mod}\, 4)$, then $n\equiv 0\, (\,{\rm
%mod}\, 4)$.
%\end{itemize}
%\end{cor}
The following result give a necessary and sufficient condition on
the existence of self-dual codes over $R_{\infty}$.
\begin{thm}
\label{th:dgself} Let $R_i$ be a finite chain ring and $K$ its
residue field with characteristic $p\ne 2$. Then then there exists a
self-dual code over $K$ if and only if there exists a self-dual code
over $R_{\infty}$.
\end{thm}
\pf From \cite[Theorem 4.7]{doughertyself} we have that if $p\neq 2$
then the lift of any self-dual codes of $K$ is a self-dual over
$R_i$. From~\cite[ Theorems 3.4]{doughertyself} we have that if
$\mathcal{C}$ is a self-dual code of length $n$ over $R_\infty$,
then $\Psi_i(\mathcal{C})$ is a self-dual code of length $n$ over
$R_i$ for all $i<\infty$.  hence the result. \qed
\section{Codes over Principal Ideal Rings}
This section considers codes over finite commutative rings which are
principal ideal rings. We give necessary and sufficient conditions
on the existence of MDS codes. It is obvious that in a finite ring
any chain of ideal is finite
 The smallest $t \ge 1$ such that
$\mathfrak a ^t=\mathfrak a ^{t+1}= \ldots$  in the chain $\mathfrak
a \supseteq \mathfrak a^2 \supseteq \mathfrak a^3 \supseteq \ldots$
is called the index of stability of $\mathfrak a$. If the ring is a
finite chain ring, $t$ is equal to the degree of nilpotency.
Furthermore for any finite commutative ring $R$ which is principal
ideal ring and with maximal ideals $\mathfrak m_1, \mathfrak
m_2,\dots, \mathfrak m_s$ with the corresponding indices of
stability  $t_1, t_2, \dots, t_s$. There exists a canonical
$R$-module isomorphism $\Psi: R^n \rightarrow \prod_{i=1}^s
(R/\mathfrak m_i^{t_i})^n$.
%principal ideal ring then it is ismorphic
%
%\begin{prop}(\cite[Proposition 2.7]{dkk}
 %\label{pirs}
%Let $R$ be a finite commutative ring. Then the following are
%equivalent
%\begin{itemize}
%\item[(i)] $R$ is a principal ideal ring.
%\item[(ii)] $R$ is isomorphic to a finite product of chain rings.
%\end{itemize}
%
%\noindent Moreover, the decomposition in (ii) is unique up to the
%order of factors. It has the form $R\cong \prod_{i=1}^s R/\mathfrak
%m_i^{t_i}$, where In this case the set $(\mathfrak m_i)_{i=1}^s$  is
%called a \textit{direct decomposition of $R$}.
%\end{prop}
%\bigskip
% Let be the canonical $R$-module isomorphism
%given by Proposition~\ref{pirs}.
For $i=1,\dots, s$, let $\mathcal{C}_i$ be a code over $ R/\mathfrak
m_i^{t_i},$ of length $n$ and let
$$\mathcal{C}=CRT(\mathcal{C}_1,\mathcal{C}_2,\dots,\mathcal{C}_s)=\Psi^{-1}(\mathcal{C}_1\times \dots \times \mathcal{C}_s)= \{\Psi^{-1} (v_1,v_2,\dots,v_s) \ |\ v_i \in
\mathcal{C}_i \}.$$ The code $C$ is called the \textit{Chinese
product of codes $\mathcal{C}_1,\mathcal{C}_2,\dots,\mathcal{C}_s$}
\cite{MDR}.

\begin{thm}\label{CRTSD}(\cite[Lemma 6.1, 6.2, Theorem 6.4]{dkk}
With the above notation, let $\mathcal{C}_1,\mathcal{C}_2, \ldots
,\mathcal{C}_s$ be codes of length $n$, with $\mathcal{C}_i$ a code
over $R_i$, and let
$\mathcal{C}=CRT(\mathcal{C}_1,\mathcal{C}_2,\dots,\mathcal{C}_s)$.
Then we have
\begin{itemize}
\item[(i)]  $|\mathcal{C}|= \prod_{i=1}^s |\mathcal{C}_i|$;
\item[(ii)] ${ rank}(\mathcal{C}) = { \max}\{{ rank}(\mathcal{C}_i)) \; |\; 1\leq i \leq s\}$;
\item[(iii)] $\mathcal{C}$ is a free code if and only if each $\mathcal{C}_i$ is a free code of
the same rank;
\item[(iv)] $d_H(CRT(\mathcal{C}_1, \mathcal{C}_2, \cdots, \mathcal{C}_s)) = {\rm min}\{d(\mathcal{C}_i))\}$;
\item[(v)] $\mathcal{C}_1,\mathcal{C}_2,\dots,\mathcal{C}_s$ are self-dual codes if and only if $\mathcal{C}$
is a self-dual code.
\end{itemize}
\end{thm}

%\noindent If $m$ is not a product of distinct primes, the situation
%is different. The following example shows that the results of
%Corollary~\ref{cor:principe} do not hold in this case.
%
%\begin{ex}
%{\em Let $R=\mathbb{Z}_4$ and
%$C=R(2,0)+R(0,2)=\{(0,0),(2,0),(0,2),(2,2)\}\subset \mathbb{Z}_4^2.$
%We have $C^{\perp}=C$, but $C$ is not free, so the corollary above
%does not hold.}
%\end{ex}

%The next example illustrates that the result of
%Corollary~\ref{cor:principe} is not true if $R$ is not a principal
%ideal ring.
Deougherty et al~\cite[Theorem 6.5]{dkk} Proved that if $R$ is a
finite principal ideal ring such that all residue fields satisfy
\begin{equation}
\label{eq:boun} |R/\mathfrak m_i|>\binom{n-1}{n-k-1}
\end{equation}
for some integers $n,k$ with $n-k-1>0$. Then there exists an MDS
$[n,k,n-k+1]$ code over $R$. In the following we will give necessary
and sufficient conditions on the existence of MDS codes over
principal ideal rings. For that we need the following results.
\begin{thm}
\label{th:MDS1} With the notation above, let
$\mathcal{C}_1,\mathcal{C}_2,\ldots,\mathcal{C}_s$ be such that each
$\mathcal{C}_i$ is a code over $R_i$, and
$\mathcal{C}=CRT(\mathcal{C}_1,\ldots, \mathcal{C}_s)$. Then the
following holds:
\begin{itemize}
\item[(i)] If $\mathcal{C}$ is an MDS code, then $\mathcal{C}$ is a free code;
\item[(ii)] $\mathcal{C}$ is an MDS code if and only if the $\mathcal{C}_i$ are MDS and have the same rank for each $i$.
\end{itemize}
\end{thm}
\pf For $(i)$, the proof is the same as for the (i) Part of
Proposition~\ref{th:5.4}. For $(ii)$, suppose $\mathcal{C}$ is MDS
and hence from (i) $\mathcal{C}$ is free. Then from Theorem
\ref{CRTSD}~(iii), the $\mathcal{C}_i$ are free and have the same
rank $k$. By Theorem~\ref{CRTSD}~(iv) and the Singleton bound, the
$\mathcal{C}_i$ are MDS. If the $\mathcal{C}_i$ are MDS and have the
same rank, then they have the same minimum distance. Then from
Theorem \ref{CRTSD}~(iii) and (iv), we have that $\mathcal{C}$ is
MDS. \qed

\noindent Now combining Theorem~\ref{th:MDS1},
Proposition~\ref{th:5.4}, and Theorem~\ref{thliftMDS}, the following
result is obtained.

\begin{thm}
\label{main3} Let $R$ be a finite principal ideal ring with
$\mathfrak m_1, \mathfrak m_2,\dots, \mathfrak m_s$ a direct
decomposition of $R$. Then an MDS code
$\mathcal{C}=CRT(\mathcal{C}_1,\ldots,\mathcal{C}_s)$ with rank $k$
exists over $R$ if and only if there exists an MDS code with the
same dimension $k$ over all of the residue fields $R/\mathfrak m_i$.
\end{thm}
Shankar~\cite{shankar} introduced the Reed-Solomon codes over
$\Z_{p_i^{e_i}}$ as the Hensel lift of Reed-Solomon codes over
fields. In the following we will define Reed-Solomn codes over
$\Z_m$.
%Let $n | p_i^l-1$ and $\xi$ be a primitive $nth$ root of
%unity such that $\overline{\xi}$ is a primitive $nth$ root of unity
%in $\ZZ{p_i}$. This gives a Reed-Solomon code over $\ZZ{p_i^{e_i}}$
%generated by $g_i(x)=(x- \xi_i)\ldots (x-\xi_i^{d-1})$.
\begin{defi} Let
$m=\prod_{i=1}^{s} p_i^{e_i}$. Then the Reed-Solomon code of minimum
distance $d$ over $\ZZ_m$ is the linear code
$\mathcal{C}=CRT(\mathcal{C}_1,\ldots,\mathcal{C}_s)$ such that for
all $1\le i\le s$, $\mathcal{C}_i$ is a Reed-Solomon code over
$\ZZ{p_i^{e_i}}$ with minimum distance $d$.
\end{defi}
\begin{prop}
With the notation above, the Reed-Solomon code defined over $\ZZ_m$
is an MDS code with minimum distance $d$.
\end{prop}
\pf From Theorem~\ref{thliftMDS} each lifted code over
$\ZZ_{p_i^{e_i}}$ is MDS with minimum distance $d$. Hence the result
follows from Theorem~\ref{main3}. \qed
\begin{ex}
\label{exe:contre} There exists an MDS code (actually an RS code)
over $\ZZ_{65}$ with length $4$ and minimum distance $d=2$. There is
also a non-trivial RS code of length $6$ over $\ZZ_{91}$ with
minimum distance $d=4$, and an MDS RS code of length 10 over
$\ZZ_{141}$.
\end{ex}

%\pf From~\cite[Theorem 3.4]{dkk}, there exists an MDS
%$[n,k,n-k+1]$ code over a finite chain ring whenever the inequality
%5(\ref{eq:boun}) is satisfied. This result can be extended to a
%finite principal ring using Theorem~\ref{main3}. \qed
\begin{rem}
The condition given by~(\ref{eq:boun}) is only a sufficient
condition on the existence of MDS codes over a principal ideal ring.
For example, the last two RS codes given in~Example~\ref{exe:contre}
are MDS but do not satisfy~(\ref{eq:boun}).
\end{rem}

\section{Constacyclic Codes over Finite Chain Rings and Formal Power Series}
In this section, constacyclic codes are considered. These codes were
first introduced as a generalization of cyclic codes over finite
fields. More recently, cyclic and negacyclic codes have been
generalized over finite chain rings and formal power
series~\cite{CS,permounth,power}. We first review and extend some
results of \cite{CS,permounth,power} to constacyclic codes. As an
application, we apply our results to construct MDS, self-dual
negacyclic codes over finite chain rings and formal power series,
and MDS codes over principal ideal rings.

Let $R$ be a finite chain ring. For a given unit $\lambda \in R$, a
code C is said to be constacyclic, or more generally,
$\lambda-$constacyclic, if $(\lambda c_{n-1}, c_0, c_1, \ldots,
c_{n-2})\in \mathcal{C}$ whenever $(c_0, c_1,\ldots, c_{n-1}) \in
C$. For example, cyclic and negacyclic codes correspond to
$\lambda=1$ and $-1$, respectively. It is well known that the
$\lambda-$constacyclic codes over a finite chain ring $R$ correspond
to ideals in $R[x]/\langle x^n-\lambda \rangle$. Recall the
definition of a formal power series over $R$ given in
(\ref{eq:infty}). It turns out that many properties of constacyclic
codes over finite chain rings also hold for constacyclic codes over
$R_{\infty}$.

Let
\[
R_\infty[x]=\{a_0+a_1x+\ldots+a_nx^n\,|\, a_i\in R_\infty, n\ge 0\}
\]
be the polynomial ring over $R_\infty$. Hence $R_\infty[x]$ is a
domain since $R_\infty$ is a domain by Lemma~\ref{lem:bou}.

The maps $\Psi^j_i$ in (\ref{eqau}) and $\Psi_i$ in (\ref{eqau1})
can be extended to maps from $R_j[x]$ to $R_i[x]$ and from
$R_\infty[x]$ to $R_i[x]$, respectively. Namely, for
$f(x)=a_0+a_1x+\ldots+a_nx^n\in R_j[x]$, we have the following maps:
\begin{eqnarray*}
\Psi_i^{j}:R_j[x]&\to& R_i[x];\\
f(x)&\mapsto&\Psi_i^{j}(f(x)),
\end{eqnarray*}
where
$\Psi_i^j(f(x))=\Psi_i^j(a_0)+\Psi_i^j(a_1)x+\ldots+\Psi_i^j(a_n)x^n$,
and
\begin{eqnarray*}
\Psi_i:R_\infty[x]&\to& R_i[x];\\
f(x)&\mapsto&\Psi_i(f(x)),
\end{eqnarray*}
where $\Psi_i(f(x))=\Psi_i(a_0)+\Psi_i(a_1)x+\ldots+\Psi_i(a_n)x^n$.
In this way, the map defined in~(\ref{eq:over}) is the same as
$\Psi_1^j$ in the finite case and $\Psi_1$ in the infinite case.

\begin{lem}
Let $\lambda_{j}$ be an arbitrary unit of $R_j$, $j\le \infty$. Then
$\Psi_i^j(\lambda_{j})$ is a unit of $R_i$.
\end{lem}
\pf Follows from~(\ref{lem:any}) and Lemma~\ref{lem:bou}. \qed

For clarity of notation, we denote $\Psi_i^j(\lambda_j)$ by
$\lambda_i$ and $\Psi_i(\lambda_{\infty})$ by $\lambda_i$ when there
is no ambiguity.

Consider now the following ring
$$
R_\infty[x]/\langle x^n-\lambda_{\infty}\rangle=\{f(x)+\langle
x^n-\lambda_{\infty}\rangle\,|\, f(x)\in R_\infty[x]\}.
$$
Since $R_\infty$ is a domain, we have that
\begin{equation}
\label{eq:inftyConsta}
 R_\infty[x]/\langle
x^n-\lambda_{\infty}\rangle=\{f(x)+\langle
x^n-\lambda_{\infty}\rangle\,|\,\mbox{where}\, \deg f(x)<n\,\,
\mbox{or}\,\, f(x)=0\}.
\end{equation}
\noindent As for the finite case a linear code $\mathcal{C}$ of
length $n$ over $R_\infty$ is called {\it a
$\lambda_{\infty}$-constacyclic code} over $R_\infty$ if it
satisfies the following implication
$$
{\bf c}=(c_0,c_1,\cdots,c_{n-1})\in
\mathcal{C}\Rightarrow(\lambda_{\infty}
c_{n-1},c_0,\cdots,c_{n-2})\in \mathcal{C}.
$$
\noindent When $\lambda_{\infty}=1$, respectively
$\lambda_{\infty}=-1$, the code $\mathcal{C}$ is called {\it
cyclic}, respectively {\it negacyclic}.

 We define the map
$P_{\lambda _{\infty}}$ as follows:
\begin{eqnarray}
\label{eq:lambda}
P_{\lambda_{\infty}}:R^n_\infty&\to& R_\infty[x]/\langle x^n-\lambda_{\infty}\rangle,\\
(a_0,a_1,\ldots,a_{n-1})&\mapsto&
a_0+a_1x+\ldots+a_{n-1}x^{n-1}+\langle x^n-\lambda_{\infty}\rangle.
\end{eqnarray}
\noindent Let $\mathcal{C}$ be an arbitrary subset of $R^n_{\infty}$
and $P_{\lambda_{\infty}}(\mathcal{C})$ the image of $\mathcal{C}$
under the map $P_{\lambda_{\infty}}$. Then we have
$$
P_{\lambda_{\infty}}(\mathcal{C})=\{c_0+c_1x+\ldots+c_{n-1}x^{n-1}+\langle
x^n-\lambda_{\infty}\rangle\,|\, (c_0,c_1,\ldots,c_{n-1})\in
\mathcal{C} \}.
$$
Hence we obtain from~(\ref{eq:inftyConsta}) and (\ref{eq:lambda})
that a linear code $\mathcal{C}$ of length $n$ over $R_\infty$ is a
$\lambda_{\infty}$-constacyclic code if and only if
$P_{\lambda_{\infty}}(\mathcal{\mathcal{C}})$ is an ideal of
$R_\infty[x]/\langle x^n-\lambda_{\infty}\rangle$.
 %\noindent This
%gives the following Lemma.
%\begin{lem}\label{cor-constacyclic}
%With the notation given above, then a linear code $\mathcal{C}$ of
%length $n$ over $R_\infty$ is a $\lambda_{\infty}-$constacyclic code
%if and only if $P_{\lambda_{\infty}}(\mathcal{C})$ is an ideal of
%$R_\infty[x]/\langle x^n-\lambda_{\infty} \rangle$.
%\end{lem}
%\begin{defi}
%Assume the notations given above. Let $f(x)\in R_\infty[x]$, if
%$\deg (f(x))>0$ and $\gcd(a_0,a_1,\cdots, a_n)=1$ then  we call
%$f(x)$ a primitive polynomial.
%\end{defi}
%
%\begin{prop}(\cite{power}
%Let $f(x)$ be a polynomial over $R_\infty$ with $\deg (f(x))>0$.
%Then $f(x)$ is a primitive polynomial if and only if
%$\Psi_i(f(x))\ne 0$ for all $i<\infty$.
%
%Furthermore for any polynomial $g(X)$ with $\deg (g(x))>0$, there
%exist a unique $s$ and a primitive polynomial $f(x)$ such that
%$g(x)=\gamma^sf(x)$.
%\end{prop}

% Now we will be considering by the factorization of
%$X^n-\lambda$ over a finite chain ring $R_i$.

For $i \leq \infty$, two polynomials $f(x),g(x)\in R_i$ are called
{\it coprime} if $\langle f(x)\rangle+\langle g(x)\rangle=R_i[x]$,
or equivalently, if there exist $u(x),v(x)\in R_i[x]$ such that
$f(x)u(x)+g(x)v(x)=1$.

Let $\mathcal{P}_i\subseteq R_i$, $i\le \infty$ be nonzero ideal.
Then  $\mathcal{P}_i$ is called a {\it prime} ideal, respectively
{\it primary} ideal if it satisfies $ab\in \mathcal{P}_i \Rightarrow
a\in \mathcal{P}_i \text{ or }b\in \mathcal{P}_i$ respectively
 $
ab\in \mathcal{P}_i  \Rightarrow a\in \mathcal{P}_i \text{ or
}b^k\in \mathcal{P}_i, $ for some positive integer $k$.

A polynomial $f(x)$ of a chain ring $R_i$ is said to be basic
irreducible if $\Psi_1^i(f)$ is irreducible in $K[x]$, where $K$ is
the residue field of $R_i$.
 A polynomial of $R_i[x]$ is called regular if it is
not a zero divisor. Hence from~(\ref{lem:any}) we have that $f\in
R_i[x]$ is regular if and only if $\Psi_1^i(f)\neq 0$.
\begin{lem}\label{hensel}
{\rm (Hensel's Lemma~\cite[Theorem XIII. 7]{Mac})} Let $i< \infty$
and $f$ be a polynomial over $R_i$. Assume $\Psi_1^i(f)=g_1g_2\ldots
g_r$ where $g_1,g_2,\ldots,g_r$ are pairwise coprime polynomials
over $K$. Then there exist pairwise coprime polynomials
$f_1,f_2,\ldots, f_r$ over $R_i$ such that $f=f_1f_2\ldots f_r$ and
$\Psi_1^i(f_j)=g_j$ for $j=1,2,\ldots,r$.
\end{lem}

\begin{lem}
\label{lem:prop2.7}(\cite[Proposition 2.7]{permounth})
 Let $f(x)$ be a monic polynomial over $R_i$, $i<
\infty$, of degree $n$ such that $\Psi_1^i(f)$ is square free. Then
$f(x)$ factors uniquely as a product of monic basic irreducible
pairwise coprime polynomials.
\end{lem}
%\begin{lem}(\cite[Exercise XIII.6]{Mac})
%\label{lem:euc} Let $f$ and $g$ be nonzero polynomials in $R_i[x]$.
%If $g$ is regular, then there exist polynomials $q$ and $r$ $\in
%R[x]$ such that $f=gq+r$.
%\end{lem}
\begin{thm}
\label{lem:aicha}Let $R_i$ be a finite chain ring with
characteristic $p$ and $\lambda_i$ a unit of $R_i$. When $(n,p)=1$,
the polynomial $x^n-\lambda_i$ factors uniquely as a product of
monic basic irreducible pairwise coprime polynomials over $R_i[x]$.
Furthermore, there is a one-to-one correspondence between the set of
basic irreducible polynomial divisors of $x^n-\lambda_i$ in $R_i[x]$
and the set of irreducible divisors of $\Psi_1^i(x^n-\lambda_i)$ in
$K$.
\end{thm}
\pf Assuming $(n,p)=1$, it must be that the componentwise reduction
modulo $\gamma$ of $x^n-\lambda_i$, which is $\Psi
_1^i(x^n-\lambda_i)$, is square free in $K[x]$. Hence by
Lemma~\ref{lem:prop2.7}, the polynomial $x^n-\lambda_i$ factors
uniquely as a product of monic basic irreducible pairwise coprime
polynomials $f_1\ldots f_s$ over $R_i[x]$. Since $K$ is a field and
hence $K[x]$ is a unique factorization domain,
$\Psi_1^i(x^n-\lambda_i)$ has a unique factorization
$h_{1}h_{2}\ldots h_{k}$ into irreducible polynomials over $K[x]$.
These are pairwise coprime since $(n,p)=1$. By Lemma~\ref{hensel},
there exist polynomials $\tilde{h_j}$ in $R_i[x]$ such that
$\Psi_1^i(\tilde{h_j})=h_j$, and $x^n-\lambda_i=\tilde{h_1}\ldots
\tilde{h_k}$. Hence the $\tilde{h_j}$ are basic irreducible. From
the fact that the decomposition of $x^n-\lambda _i$ over $R_i[x]$ is
unique, we obtain that $\tilde{h_j}=f_j$ and $k=s$. \qed

In the following we focus on constacyclic codes over $R_{\infty}$
and the projections of these codes.
% constacyclic codes over these rings.
%
Let
\begin{eqnarray}
\label{homo}\Psi_i:R_\infty[x]/\langle x^n-
\lambda_{\infty}\rangle&\to& R_i[x]/\langle
x^n-\lambda _i\rangle\\
f(x)&\mapsto& \Psi_i(f(x)).
\end{eqnarray}
The map of (\ref{homo}) is a ring homomorphism. Thus if $I$ is an
ideal of $R_\infty[x]/\langle x^n-\lambda_{\infty}\rangle$, then
$\Psi_i(I)$ is an ideal of $R_i[x]/\langle x^n-\lambda_i\rangle$.
This gives the following commutative diagram:
$$
\begin{CD}
 R_\infty^n@>P_{\lambda_{\infty}}>>R_\infty[x]/\langle x^n-\lambda_{\infty}\rangle\\
@V \Psi_i VV
 @VV \Psi_i V \\
 R_i^n@>P_{\lambda_i}>>R_i[x]/\langle x^n-\lambda_i)\rangle.
\end{CD}
$$
Hence we have the following theorem.
\begin{thm}\label{projection-1}
The projection code $\Psi_i(\mathcal{C})$ of a
$\lambda_{\infty}-$constacyclic code $\mathcal{C}$ of $R_\infty$ is
a $\lambda_i-$constacyclic code over $R_i$ for all $i<\infty$.
\end{thm}
\pf Assume that $C$ is a $\lambda_{\infty}-$constacyclic code over
$R_\infty$. Then $P_{\lambda_{\infty}}(\mathcal{C})$ is an ideal of
$R_\infty[x]/\langle x^n-\lambda_{\infty}\rangle$. By the
homomorphism in (\ref{homo}) and the commutative diagram above,
$\Psi_i(P_{\lambda_{\infty}}(\mathcal{C}))=P_{\lambda_i}(\Psi_i(\mathcal{C}))$
is an ideal of $R_i[x]/\langle x^n-\lambda_i) \rangle$. This implies
that $\Psi_i(\mathcal{C})$ is a $\lambda_i-$constacyclic code over
$R_i$ for all $i<\infty$.\qed

%\begin{thm}
%Assume the notations given above. Let $i<j<\infty$ be two
%integers. Let $C$ be a code over $R_j$. If $\Psi_i^j(C)$ is a
%cyclic code over $R_i$ then $C$ is a cyclic code over $R_j$.
%\end{thm}
\begin{lem}
\label{lem:domain3} Let $\mathcal{C}$ be a
$\lambda_{j}-$constacyclic code over $R_j$, $j\le \infty$, and
$\mathcal{C}^{\bot}$ the dual code of $\mathcal{C}$. Then the code
$\mathcal{C}^{\bot}$ is a $\lambda_{j}^{-1}-$constacyclic code over
$R_j$.
\end{lem}
\pf We have that $\lambda_j$, $j\le \infty$, is a unit. Furthermore
since $i<\infty$ we have that $R_i$ is a finite chain ring. From
Lemma \ref{lem:bou} $R_\infty$ is a principal ideal domain. Hence
the ideals of $R_j$ are principal. Hence the result follows by a
proof similar to that for constacyclic codes over a finite field.
\qed
\begin{thm}
Let $\mathcal{C}$ be a $\lambda_{\infty}-$constacyclic code over
$R_\infty$ and $\mathcal{C}^{\bot}$ the dual code of $\mathcal{C}$.
Then the code $\Psi_i(\mathcal{C}^{\bot})$ is a
$\lambda_i^{-1}-$constacyclic code, and if
$\mathcal{(C^{\bot})^{\bot}}=\mathcal{C}$ then
$\Psi_i(\mathcal{C}^{\bot})=\Psi_i(\mathcal{C})^{\bot}$ for all
$i<\infty$.
\end{thm}
\pf From Lemma~\ref{lem:domain3} we have that $C^{\bot}$ is a
$\lambda_{\infty}^{-1}-$constacyclic code over $R_{\infty}$. Hence
from Theorem~\ref{projection-1}, the code
$\Psi_i(\mathcal{C}^{\bot})$ is a
$\Psi_i(\lambda_{\infty}^{-1})$-constacyclic code for all
$i<\infty$. Then since $\Psi_i$ is a ring homomorphism and the rings
are with unity, we have
$\Psi_i(\lambda_{\infty}^{-1})=\lambda_{i}^{-1}$. Hence the result
follows. Now we prove that
$\Psi_i(\mathcal{C}^{\bot})=\Psi_i(\mathcal{C})^{\bot}$ for all
$i<\infty$.

Let $v\in \Psi_i(\mathcal{C}^{\bot})$ and let $w$ be an arbitrary
element of $\Psi_i(\mathcal{C})$. Then there exist $v'\in
\mathcal{C}^{\bot}$ and $w'\in \mathcal{C}$ such that $v=\Psi_i(v')$
and $w=\Psi_i(w')$. We have that $v\cdot w=\Psi_i(v')\cdot
\Psi_i(w')=\Psi_i(v'\cdot w')=\Psi_i(0)=0.$ This implies that
$\Psi_i(\mathcal{C}^{\bot})\subseteq (\Psi_i(\mathcal{C}))^{\bot}$.
By Lemma~\ref{type}, $\mathcal{C}^{\perp}$ has type $1^{n-k}$. Since
$\mathcal{C}=(\mathcal{C}^{\perp})^{\perp}$, by Lemma~\ref{type},
this implies that $\mathcal{C}$ has type $1^k$. Hence
$\Psi_i(\mathcal{C}^{\bot})$ has type $1^{n-k}$ and
$(\Psi_i(\mathcal{C}))^{\bot}$ has type $1^{n-k}$. It was proven
already that $\Psi_i(\mathcal{C}^{\bot})\subseteq
(\Psi_i(\mathcal{C}))^{\bot}$. Hence
$(\Psi_i(\mathcal{C}))^{\bot}=\Psi_i(\mathcal{C}^{\bot})$.\qed

\begin{lem}\label{gamma}
Assume the notation given above and let $\mathcal{P}_i$ be an
arbitrary prime ideal of $R_i[x]/\langle x^n-\lambda_i \rangle$, for
$i < \infty$. Then we have $\gamma\in \mathcal{P}_i$.
\end{lem}
\pf Since $\mathcal{P}_i$ is an ideal and the nilpotency index of
$\gamma$ is $i$, we have that $\gamma^i=0\in \mathcal{P}_i$. As
$\mathcal{P}_i$ is prime, either $\gamma^{i-1}\in \mathcal{P}_i$ or
$\gamma \in \mathcal{P}_i$. Assume $\gamma\not\in \mathcal{P}_i$,
then $\gamma^{i-1}\in \mathcal{P}_i$. Again since $\mathcal{P}_i$ is
prime, and $\gamma\not\in \mathcal{P}_i$, then $\gamma^{i-2}\in
\mathcal{P}_i$. Continuing this process we obtain that $\gamma^2 \in
\mathcal{P}_i$, and hence $\gamma \in \mathcal{P}_i$, which is a
contradiction.\qed

\begin{thm}
\label{th:Prime} Assume the notation given above. Then the prime
ideals in $R_i[x]/\langle x^n-\lambda_i \rangle$ are $\langle
\pi_i(x), \gamma\rangle$, where $\pi_i(x)$ is a monic basic
irreducible polynomial divisor of $x^n-\lambda_i$ over $R_i$. If
$i=\infty$, then the ideals $\langle \pi_i(x)\rangle$, where $i\ge
1, i\in\mathbb{N}$ are also prime ideals of $R_{\infty}[x]/\langle
x^n-\lambda_{\infty} \rangle$.
\end{thm}
\pf For the finite case, let $\mathcal{P}_i$ be an arbitrary prime
ideal in $R_i[x]/\langle x^n-\lambda_i\rangle$. Since $\Psi_1^i$ is
a ring homomorphism, $\Psi_1^i(\mathcal{P}_i)$ is also a prime ideal
in $K[x]/\langle x^n-\lambda_1 \rangle$. Since $K$ is a field, any
prime ideal in $K[x]/\langle x^n-\lambda_1 \rangle$ over $K$ is of
the form $\langle \pi_1(x)\rangle$~\cite[Theorem 3.10]{Mac2}, where
$\pi_1(x)$ is a monic irreducible divisor of $x^n-\lambda_1$ over
$K$. Hence $\Psi^i_1(\mathcal{P}_i)=\langle \pi_1(x)\rangle$, and
$\pi_1(x)\in\langle \pi_1(x)\rangle=\Psi^i_1(\mathcal{P}_i)$. By
Lemma~\ref{hensel}, there exists $\pi_i(x)\in \mathcal{P}_i$ such
that $\Psi^i_1(\pi_i(x))=\pi_1(x)$, where $\pi_i(x)$ is a monic
basic irreducible divisor of $x^n-\lambda_i$ over $R_i$. Since
$i<\infty$, by Lemma~\ref{gamma}, we have that $\gamma\in
\mathcal{P}_i$. This implies that $\langle \pi_i(x),
\gamma\rangle\subseteq \mathcal{P}_i$. We have that $(R_i[x]/\langle
x^n-\lambda_i \rangle)/\langle \pi_i(x), \gamma\rangle$ is a field,
so $\langle \pi_i(x), \gamma\rangle$ is maximal, and thus
$\mathcal{P}_i=\langle \pi_i(x), \gamma\rangle$.

For $i=\infty$ and $\gamma\not\in \mathcal{P}_i$, the only other
possibility is $\mathcal{P}_i=\langle \pi_{i}(x)\rangle$. \qed

\begin{thm}
\label{th:idem} Every prime ideal $\mathcal{P}_i=\langle \pi_i(x),
\gamma\rangle$ in $R_i[x]/\langle x^n-\lambda_i\rangle$ contains an
idempotent $e_i(x)$ with $e_i(x)^2=e_i(x)$, and
$\mathcal{P}_i=\langle e_i(x), \gamma\rangle$. Furthermore, if
$i=\infty$, then every prime ideal $\mathcal{P}_i=\langle
\pi_{\infty}(x)\rangle$ of $R_\infty[x]/\langle x^n-\lambda_{\infty}
\rangle$ has an idempotent generator.
\end{thm}
\pf  We establish the first assertion by induction. Let $K$ be the
residue field of characteristic $p$ of $R_i$. Then since we can
apply the Euclidean algorithm over $K[x]$, by a proof similar to
that for the cyclic case in~\cite[Ch. 8 Theorem 1]{MS}, we have that
every ideal $\mathcal{P}_1$ in $K[x]/\langle x^n-\lambda_1\rangle$
contains an idempotent $e_1$ such that $\mathcal{P}_1=\langle e_1
\rangle$. Let $\langle\Psi_l^i(\pi_i(x)),\gamma\rangle$ be the
projection of $\mathcal{P}_i=\langle(\pi_i(x)),\gamma\rangle$ onto
$R_l[x]/\langle x^n-\lambda_l\rangle$. Suppose $e_l(x)\in
\langle\Psi_l^i(\pi_i(x)),\gamma\rangle$ is an idempotent element
with $\langle e_l(x), \gamma\rangle=\langle \Psi_l^i(\pi_i(x)),
\gamma\rangle$. Then we have that $e_l^2(x)=e_l(x)+\gamma^lh(x)$ in
$R_{l+1}[x]/\langle x^n-\lambda_{l+1}\rangle$ for some $h(x) \in
R_{l+1}[x]/\langle x^n-\lambda_{l+1}\rangle$. In the following, we
show that $e_{l+1}(x)=e_l(x)+\gamma^l\theta(x)$ is an idempotent
element by choosing a suitable $\theta(x)$. We have that
\begin{eqnarray*}
e^2_{l+1}(x)&\equiv&(e_l(x)+\gamma^l\theta(x))^2=e^2_l(x)+2\gamma^l\theta(x)e_l(x)\,\pmod{\gamma^{l+1}}\\
&\equiv&e_l(x)+\gamma^lh(x)+2\gamma^l\theta(x)e_l(x)\,\pmod{\gamma^{l+1}}\\
&\equiv&e_{l+1}(x)-\gamma^l\theta(x)+\gamma^lh(x)+2\gamma^l\theta(x)e_l(x)\,\pmod{\gamma^{l+1}}\\
&\equiv&e_{l+1}(x)+\gamma^l(h(x)-\theta(x)(1-2e_l(x)))\,\pmod{\gamma^{l+1}}.
\end{eqnarray*}
If $p=2$, we can choose $\theta(x)=h(x)$, and $e_{l+1}(x)$ is an
idempotent element. If $p\ne 2$, then
$(1-2e_l(x))^2=1+4\gamma^lh(x)$. This gives that $(1-2e_l(x))$ is a
unit. Then by choosing $\theta(x)=h(x)(1-2e_l(x))^{-1}$, we get that
$e_{l+1}(x)$ is an idempotent element in $R_{l+1}[x]/\langle
x^n-\lambda_{l+1}\rangle$, and then $\langle e_{l+1}(x),
\gamma\rangle = \langle \pi_{l+1}(x), \gamma \rangle$.

Since $\pi_{\infty}(x)$ and $(x^n-\lambda_{\infty})/\pi_{\infty}(x)$
are relatively prime, there exist $h(x),h'(x)\in R_\infty[x]$ such
that
$$h(x)\pi_{\infty}(x)+h'(x)\cdot((x^n-\lambda_{\infty})/\pi_{\infty}(x))=1.$$
This means that
$$
(h(x)\pi_{\infty}(x))^2=h(x)\pi_{\infty}(x)-h'(x)h(x)\cdot(x^n-\lambda_{\infty}),
$$
and hence
$$
(h(x)\pi_{\infty}(x))^2\equiv
h(x)\pi_{\infty}(x)\,\pmod{x^n-\lambda_{\infty}}.
$$
Then $h(x)\pi_{\infty}(x)$ is an idempotent element in
$R_\infty[x]/\langle x^n-\lambda_{\infty}\rangle$.\qed
\begin{thm}\label{decm}
Assume the notation given above. Then for $i \le \infty$, the
primary ideals in $R_i[x]/\langle x^n-\lambda_i \rangle$ are
$\langle 0 \rangle $, $\langle 1 \rangle$, $\langle
\pi_i(x)\rangle$, $\langle \pi_i(x),\gamma^l\rangle$, where $
\pi_i(x)$ is a basic irreducible divisor of $x^n-\lambda_i$ over
$R_i$ and $1\le l<i$.
\end{thm}
\pf  Let $\mathcal{P}_i$ be a prime ideal of $R_i[x]/\langle
x^n-\lambda_i \rangle$. Hence by Theorem~\ref{th:Prime},
$\mathcal{P}_i=\langle \pi_i(x),\gamma\rangle$ and if $i=\infty$,
there is another case $\mathcal{P}_i=\langle \pi_i(x)\rangle$. It is
obvious that these prime ideals are primary. Then the first class of
primary ideals of $R_i[x]/\langle x^n-\lambda_i \rangle$ is the
class of prime ideals given in Theorem~\ref{th:Prime}. From the fact
that $\langle \gamma\rangle$ is maximal in $R_i$,
$\mathcal{P}_i=\langle \pi_i(x),\gamma\rangle$ is maximal in
$R_i[x]/\langle x^n-\lambda_i \rangle$, but $\mathcal{P}_i=\langle
\pi_i(x)\rangle$ is not maximal. By \cite[Corollary 2, p 153]{Z-S},
we have that the powers of the maximal ideals are primary ideals.
Let $\mathcal{Q}_i$ be a primary ideal associated with the prime
ideal $\mathcal{P}_i=\langle \pi_i(x),\gamma\rangle$. Then
by~\cite[Ex. 2, p. 200]{Z-S}, there is an integer $k$ such that
$\mathcal{P}_i^k \subset \mathcal{Q}_i \subset \mathcal{P}_i$. From
this, we obtain $\mathcal{Q}_i =\mathcal{P}_i^l$, from some $l$.
Hence the primary ideals of $R_i[x]/\langle x^n-\lambda_i \rangle$
are $(\langle \pi_i(x),\gamma\rangle)^l$ and $\langle
\pi_i(x)\rangle.$ From~Theorem~\ref{th:idem}, we have that
$\mathcal{P}_i=\langle\pi_i(x),\gamma\rangle=\langle
e_i(x),\gamma\rangle$, and $e_i(x)$ is an idempotent of
$R_i[x]/\langle x^n-\lambda_i \rangle$. Hence
$\mathcal{P}_i^l=(\langle\pi_i(x),\gamma\rangle)^l=(\langle
e_i(x),\gamma\rangle)^l$. Let $a\in (\langle
e_i(x),\gamma\rangle)^l$, then there exist $g_{t,i}(x),
h_{t,i}(x)\in R_i[x]$, such that
$a=\prod_{t=1}^l(e_i(x)g_{t,i}(x)+\gamma h_{t,i}(x))$. Since
$e_i(x)^2=e_i(x)$, then $a= e_i(x)G_i(x)+\gamma^l H_i(x)$ for some
$G_i(x), H_i(x) \in R_i[x]$. Hence the non trivial primary ideals of
$R_i[x]/\langle x^n-\lambda_i \rangle$ are $\langle \pi_i\rangle$
and $\langle \pi_i, \gamma^l \rangle$. \qed

\begin{thm}\label{decom}
Let $\pi^l_i(x), 1\le l\le b, i\in\NN$, denote the distinct monic
irreducible divisors of $x^n-\lambda_i$ over $R_i$, with $i\le
\infty$.
% and $\NN$ is the set of natural numbers.
Then any ideal in $R_i[x]/\langle x^n-\lambda_i\rangle$ can be
written in a unique way as follows
\begin{equation}
\label{eq:prod} I=\prod_{l=1}^b\langle
\pi_i^l(x),\gamma^{m_l}\rangle,
\end{equation}
where $0\le m_l\le i$. In particular, if $i$ is finite, then there
are $(i+1)^b$ distinct ideals.
\end{thm}
\pf Since $R_i[x]/\langle x^n-\lambda_i\rangle$ is Noetherian, from
the Lasker-Noether decomposition Theorem \cite[p. 209]{Z-S} any
ideal in $R_i[x]/\langle x^n-\lambda_i\rangle$ has a representation
as a product of primary ideals. From Theorem~\ref{decm}, we have
that the primary ideals of $R_i[x]/\langle x^n-\lambda_i\rangle$ are
$\langle {\pi^l}_i(x),\gamma^{m_l}\rangle$. Hence the result
follows. In addition, if $i$ is finite then there are $(i+1)^b$
distinct ideals in $R_i$. \qed

The following lemma is a generalization of Hensel's Lemma.
\begin{thm}
\label{thliftMDS1} Let $\lambda_i$ be a unit in a chain ring, $i \le
\infty$. If $h_1(x)\in K[x]$ is a monic irreducible divisor of
$x^n-\lambda_1$ such that $K$ is the residue field of $R_i$, then
there is a unique monic irreducible polynomial $h_i$ which divides
$(\Psi_1^i)^{-1}(x^n-\lambda _1)$ over $R_i$ and is congruent to
$h_1(x) \pmod \gamma.$
\end{thm}
 \pf Let $f(x)$ be the lift of $h_1(x)$ over $R_{\infty}$.
 If $f(x)$ is reducible over $R_\infty$ then there exist
polynomials  $g(x), h(x)$ such that $f(x)=g(x)h(x)$ and $0 < \deg
(g(x)),\deg (h(x))<\deg (f(x))$. This implies that
$$
\Psi_1(f(x))=\Psi_1(g(x)h(x))=\Psi_1(g(x))\Psi_1(h(x))=h_1(x).
$$
Since $f(x)$ is monic, we have that $0<\deg (\Psi_1(g(x))),\deg
(\Psi_1(h(x)))<\deg (\Psi_1(f(x)))=\deg (h_1(x))$. This is a
contradiction. Since $f(x)$ is irreducible, $\langle f(x) \rangle $
is a prime ideal of $R_{\infty}$. In addition, $f(x)$ must be a
divisor of $\Psi_1^{-1}(x^n-\lambda _1)$, otherwise $\Psi_1(f)=h_1$
is not a divisor of $x^n-\lambda_1$. Since $\langle f(x) \rangle $
is maximal in $R_\infty[x]/\langle
\Psi_1^{-1}(x^n-\lambda_1)\rangle$, $f(x)$ is unique. If $i <
\infty$ the result follows from Theorem~\ref{lem:aicha}.
 \qed

%Theorem~\ref{} these
%ideal are uniques, hence
%Since by Lemma~\ref{lem:bou}we have $R_infty$ is a principal ideal
%domain, then $\langle f(x) \rangle $ is a maximal ideal~\cite[p
%244]{Z-S} we have that the ideal is maximal and hence unique. \qed

\begin{thm}\label{cyclic-sructure}
Let $R_i$ be a chain ring $i\le \infty$, and  $C$ be a constacyclic
code of length $n$ over $R_i[x]/\langle x^n-\lambda_i\rangle$.
\begin{itemize}
\item[(i)] If $i<\infty$, then $C$ is  equal to
\begin{equation}
\langle g_0(x),\gamma g_1(x),\cdots,\gamma^{i-1}g_{i-1}(x)\rangle,
\end{equation}
where the $g_l(x)$ are divisors of $x^n-\lambda _i$ and
$g_{i-1}(x)\,|\,\cdots|\,g_1(x)\,|\,g_0(x)$.
\item[(ii)] If  $i=\infty$, then $C$ is equal to
\begin{equation}
\label{sum} \langle
\gamma^{t_0}g_0(x),\gamma^{t_1}g_1(x),\cdots,\gamma^{t_{b-1}}g_{l-1}(x)\rangle,
\end{equation}
where $0\le t_0<t_1<\cdots<t_{l-1}$ for some $l$ and
$g_{l-1}(x)\,|\,\cdots|\,g_1(x)\,|\,g_0(x)$.
%In particular, each
%ideal of $R_\infty[x]/\langle x^n-1\rangle$ is principal.
\end{itemize}
\end{thm}
\pf The results follows by expanding the product in
Theorem~\ref{decom} from Theorem~\ref{decm} \qed

\begin{thm}
\label{th:prince} Let $\mathcal{C}$ be a constacyclic code over
$R_i[x]$. If $i < \infty$, then there exists a unique family of
pairwise coprime polynomials $F_0, \ldots, F_i$ in $R_i[x]$ such
that $F_0 \ldots F_i=x^n-\lambda _i$ and $\mathcal{C}=\langle\hat{
F}_1+\gamma \hat{F}_2+ \ldots + \gamma^i \hat{F}_i \rangle $, where
$\hat{F}_j=\frac{x^n-1}{F_j}$, for $0<j<i$. Moreover
\begin{equation}
\label{car:cons} |\mathcal{C}|=|K|^{\sum_{j=0}^{i-1}(i-j)\deg
F_{j+1}}.
\end{equation}
\end{thm}
\pf The proof is similar to that for the cyclic case \cite[Theorem
3.8]{permounth}.
\begin{cor} With the above notation,
for $i \le \infty$, every ideal in $R_i[x]/\langle x^n -\lambda_i
\rangle$ is principal.
\end{cor}
\pf For $i<\infty$, the result is given by Theorem~\ref{th:prince}.

For $i=\infty$, let $I$ be an ideal in $R_{\infty}[x]/\langle
x^n-\lambda_{\infty} \rangle$, with $\lambda_{\infty}$ a unit in
$R_{\infty}$. Then $\Psi_j(I)$ is a principal ideal $\langle g_j
\rangle $ of $R_j[x]/\langle x^n-\lambda_j \rangle$ for all
$0<j<\infty$ from the first case. Using~(\ref{eq:com}), we can
define a $\gamma -$adic metric, since $R$ is finite hence by
Tychonoff's theorem~\cite{willard} $R_{\infty}$ is compact and then
$R_{\infty}[x]/\langle x^n -\lambda_{\infty} \rangle$ is also
compact with respect to this metric. Hence the sequence $\{g_j\}$
has a subsequence which converges to a limit $g$, which gives the
result. \qed

Now we consider free constacyclic codes as free linear codes over
the finite chain rings defined in Section 2.
\begin{thm}
\label{th:free} Let $\mathcal{C}$ be a $\lambda_i-$constacyclic code
of length $n$ over a finite chain ring $R_i$ with characteristic $p$
such that $(p,n)=1$. Then $\mathcal{C}$ is a free constacyclic code
with rank $k$ if and only if there is a polynomial $f(x)$ such that
$f(x)|(x^n-\lambda_i)$ which generates $\mathcal{C}$. In this case,
we have $k=n-deg(f)$.
\end{thm}
\pf Let $f(x)$ be a polynomial of degree $r$ such that
$f(x)|(x^n-\lambda _i)$, and $\mathcal{C}=\langle f(x)\rangle $ be
the constacyclic code generated by $f(x)$ such that $\deg f=r$.
Assume that $f_0$ and $f_r$ are the constant and leading
coefficients of $f$, respectively. Then $f_0$ and $f_r$ are units in
$R_i$, since $x^n-\lambda_i $ is monic and $\lambda_i$ is a unit.
Let $B=\{f(x), xf(x), \ldots, x^{n-r-1}f(x)\}$. We will prove that
$B$ is a basis for $\mathcal{C}$. First, it is established that the
vectors are independent. Suppose
\begin{equation}
\label{eq:set} \alpha_0f(x)+\ldots +\alpha_{n-r-1}x^{n-r-1}f(x)=0,
\end{equation}
where $\alpha_0,\ldots,\alpha_{n-r-1} \in R$. By comparing
coefficients, we have $\alpha_0 f_0=0$, but since we noticed that
$f_0$ is a unit then we obtain $\alpha_0=0$. Hence (\ref{eq:set})
becomes
\begin{equation}
 \alpha_1 f(x)+\ldots +\alpha_{n-r-1}x^{n-r-1}f(x)=0.
\end{equation}
Again by comparing the coefficients we obtain $\alpha_1 f_0=0$. This
also gives $\alpha_1=0$. We finally obtain $\alpha_0=\ldots
=\alpha_{n-r-1}=0$, and therefore the vectors of $B$ are linearly
independent.

Now we prove that $B$ spans $\mathcal{C}$. Let $c(x)\in \langle f(x)
\rangle$. Then there is a polynomial $g(x)\in R[x]$ such that
$c(x)=g(x)f(x)$, where $\deg g \leq n-1$. If $\deg g(x)\leq n-r-1,$
then $c(x)\in span(B)$. Otherwise, since $f$ is a regular polynomial
(divisor of $x^n-\lambda_i$ with $(n,p)=1$), then by~\cite[Exercise
XIII.6]{Mac} there are polynomials $p(x),q(x)$ such that
\begin{equation}
\label{eq:div} g(x)=\frac{x^n-\lambda_i}{f(x)}p(x)+q(x),
\end{equation}
where $\deg q(x) \leq n-r-1$. Now multiplying (\ref{eq:div}) by
$g(x)$ gives
\begin{equation}
f(x)g(x)=f(x)q(x).
\end{equation}
Hence $c(x) \in span(B)$, which gives that the code $\mathcal{C}$ is
a free $R$ module.

In order to prove the converse, suppose that $\mathcal{C}=\langle
\hat{F}_1+\gamma \hat{F}_2+\ldots+\gamma ^{i-1}\hat{F}_{i} \rangle$
is a free code of rank $k$. Hence $\mathcal{C}$ has a basis of
cardinality $k$. Consider now the polynomial $F=\hat{F}_1+\gamma
\hat{F}_2+\ldots+\gamma ^{i-1}\hat{F}_{i}$. We prove that $ \deg F=
n-k$. Let $s=n-\deg F$ and
$\Psi_1^i(\mathcal{C})=Tor_0(\mathcal{C})$. Then from
(\ref{eq:tor1}) we have that $|\Psi_1^i(\mathcal{C})|=p^{rk}$. On
the other hand, the image $\Psi_1^i(F)$ of $F$ modulo $\gamma$ is a
generator of $\Psi_1^i(\mathcal{C})$. This implies that $x^s
\Psi_1^i(F(x))$, and any power $x^l\Psi_1^i(F(x))$, $l \geq s$, can
be written as a linear combination of $\{\Psi_1^i(F(x)),
x\Psi_1^i(F(x)),\ldots, x^{s-1}\Psi_1^i(F(x))\}$. This set is also
independent and hence is a basis of $\Psi_1^i(\mathcal{C})$,
%since la base est de cardinalite k
 which gives that $|\Psi_1^i(\mathcal{C})|=p^{rs}$, so that $k=s=n-\deg F$.
By equating (\ref{eq:car}) and (\ref{car:cons}), we have that each
$k_j=\deg F_{j_l+1}$ for some $j_l \in \{0,i-1\}$. Hence from
(\ref{eq:car1}) we have $k=\sum k_j= \sum \deg {F}_{j+1}=n=n-\deg
F$, which is possible if and only if $k_{j}=0$ for $i >0$. Hence
$k=k_0=n-\deg F$. \qed

\begin{thm}
\label{th:num} Let $R_i$, $i \leq \infty$, be a chain ring and $K$
its residue field. Let $\mathcal{C}$ be a $\lambda_1-$constacyclic
MDS code of length $n$ over $K$. Then there is a unique MDS code
$\tilde{\mathcal{C}}$ over $R_i$ which is the lifted code of
$\mathcal{C}$ over $R_i$. The code $\tilde{\mathcal{C}}$ is a free
constacyclic code with generator polynomial $(\Psi_1^i)^{-1}(g)$, a
monic polynomial divisor of $(\Psi_1^i)^{-1}(x^n-\lambda_1)$, and
$d_H(\tilde{\mathcal{C}})=d_H(\mathcal{C})$.
\end{thm}
\pf By Theorem~\ref{thliftMDS}, the code $\tilde{\mathcal{C}}$ is
MDS. Hence from Theorem~\ref{th:5.4} we have that the code is a free
code, and from Theorem~\ref{th:free} $\tilde{\mathcal{C}}$ is
generated by $(\Psi_1^i)^{-1}(g)$ a divisor of
$(\Psi_1^i)^{-1}(x^n-\lambda_1)$. From Theorem~\ref{thliftMDS1}, we
have that $(\Psi_1)^{-1}(g)$ is monic and unique.
%Since $(n,p)=1$, hence $g(x)$ is square free and then  by Lemma~\ref{lem:aicha} factorize
%uniquely as a product of basic reducible polynomial.
Furthermore, Theorem~\ref{thliftMDS} and Lemma~\ref{th:min} give
that $d_H(\tilde{\mathcal{C}})=d_H(\mathcal{C})$. \qed

\begin{thm}
Let $R_i$ be a finite chain ring with nilpotence index $i$. Let
$C_{rp}(n)$ be the number of $rp$-cyclotomic classes modulo $n$ with
$(n,pr)=1$. Further, let $\lambda _i$ be a unit in $R_i$ such that
$\lambda _i^r=1$. Then the following holds:
\begin{itemize}
 \item[(i)]the number of constacyclic codes over $R_i$ is
equal to $(i+1)^{C_{rp}(n)},$
\item[(ii)]the number of free constacyclic codes
over $R_i$ is equal to $2^{C_{rp}(n)}$.
\end{itemize}
\end{thm}
\pf It follows from Corollary \ref{decom} that the number of
constacyclic codes over $R_i$ is equal to
\begin{equation}
\label{eq:num} (i+1)^s.
\end{equation}
By Theorem~\ref{lem:aicha}, this number is equal to the number of
irreducible polynomials in the factorization of $x^n-\lambda_1$ over
$K$, which is also equal to the number of $rp$-cyclotomic classes
modulo $n$. This proves $(i)$.

Part $(ii)$ follows from~Theorem~\ref{th:num} and the fact that the
number of divisors of $x^n-\lambda _1$ over an extension of $K$ of
degree $r$ is equal to the number of $rp$-cyclotomic classes modulo
$n$~\cite{huffman03}. \qed

\section{MDS Self-Dual Codes from Cyclic and Negacylic Codes}

The following result was given in~\cite[Theorem 11, 12]{guenda11}.
\begin{lem}
\label{lem:kenza2} Let $n$ be an even integer and $q$ an odd prime
power. Then there exist MDS negacyclic codes over $\F_q$ which are
self-dual codes in the following cases:
\begin{enumerate}
 \item[(i)] $n=2n'$ with $n'$ odd $q\equiv 1 \pmod 4$, and $n|q+1$;
\item[(ii)] $n=2^an'$ with $n'$ odd, $q\equiv 1 \pmod 2^{a+1}n'$, and $n|q-1$.
\end{enumerate}
\end{lem}
Let $q=p$ be an odd prime and $n$ an even integer as in
Lemma~\ref{lem:kenza2}. Then there exists a negacyclic MDS self-dual
code of length $n$ over $\ZZ_p$. Assume now that $K$ is a finite
field such that $|K|=p$. Then codes exist which are isomorphic to
those given by~Lemma~\ref{lem:kenza2}. From Theorem~\ref{th:num},
these codes are lifted to MDS negacyclic codes over~$R_i$ if
$i<\infty$. For $i=\infty$, the lifted codes are also MDS by
Theorem~\ref{thliftMDS} and negacyclic by Theorem~\ref{thliftMDS1}.
These lifted codes are also self-dual from Theorem~\ref{th:dgself}.
Hence we have the following result.
\begin{thm}
\label{lem:neg} Let $n$ be an even integer and $p$ an odd prime such
that $(n,p)=1$. Let $R_i$, $i\le \infty$, be a chain ring with
residue field $K$ such that $|K|=p$. Then there exists an infinite
family of negacyclic codes over $R_i$ which are MDS and self-dual in
the following cases:
\begin{enumerate}
 \item[(i)] $n=2n'$ with $n'$ odd, $p\equiv 1 \pmod 4$, and $n|p+1$;
\item[(ii)] $n=2^an'$ with $n'$ odd, $p\equiv 1 \pmod 2^{a+1}n'$, and $n|p-1$.
\end{enumerate}
\end{thm}
%Another natural question that arises is whether or not the
%self-duality is conserved between the codes $C_i$ and the code
%$C=CRT(C_1, \ldots, C_k)$. The following lemma gives a positive
%answer.
%\begin{lem}\cite[Theorem 6.4]{dkk}
%\label{lem:self2}  The codes $C_i$ are self-dual if and only if
%$C=CRT(C_1,\ldots,C_k)$ is a self-dual code over $\ZZ_m$.
%\end{lem}
In~\cite[Theorem 7]{guenda11}, the following existence results for
MDS self-dual codes over $\FF_q$ were given.
\begin{lem}
\label{lem:kenza} There exist $[n+1,\frac{n+1}{2},\frac{n+3}{2}]$
MDS self-dual codes which are extended odd-like duadic codes
$\widetilde{D_i}$ in the following cases:
\begin{enumerate}
\item[(i)] $q=r^t$ with $r\equiv 3 \pmod 4$, $t$ odd and $n=p^m$, with
$p$ a prime such that $p \equiv 3\pmod 4$ and $m$ odd;
\item[(ii)] $q=r^t$ with $t$ odd, $p$ a prime such that $r \equiv p \equiv 1 \pmod 4$ and
$n=p^m$.
\end{enumerate}
\end{lem}
Now we prove the existence of an infinite family of MDS self-dual
codes over $\ZZ_m$.
\begin{thm}
\label{th:ru1} Let $n$ be an even integer,
$m=\prod_{i=1}^sp_i^{e_i}$, and $p_i$ such that $n$ divides $p_i-1$
for all $1\leq i\leq s$. Then there exist $MDS$ self-dual codes over
$\Z_m$ derived from the extended duadic codes over $\ZZ_{p_i}$ in
the following cases
\begin{enumerate}
\item[(i)] $n \equiv 0 \pmod 4$ and $p_i\equiv 3 \pmod 4$, for all $1\leq i \leq s$;
\item[(ii)] $n\equiv 2 \mod 4$ and $ p_i\equiv 1 \pmod 4$, for all $1\leq i \leq s$.
\end{enumerate}
\end{thm}
\pf From the above conditions and~Lemma~\ref{lem:kenza}, we have the
existence of MDS self-dual codes over $\ZZ_{p_i}$ for all $1\leq i
\leq s$. Hence from Theorem~\ref{thliftMDS}, these MDS codes over
$\ZZ_{p_i}$ can be lifted to MDS codes over $\ZZ_{p_i^j}, j>1$.
Theorem~\ref{th:MDS1} proves that we have MDS codes over $\Z_m$.
Furthermore, from Theorems~\ref{CRTSD} and \ref{th:dgself}, they are
self-dual. \qed

In Table 1, we give examples of self-dual MDS codes over $\ZZ_m$
obtained using the results above. Codes over fields
from~\cite{bet,guenda11,elias} were also used to obtain codes over
$\ZZ_m$.
\begin{table}
\label{tab:self} \caption{Some Self-dual MDS Codes of Length $n$
over $\ZZ_m$}
\begin{center}
\begin{tabular}{|c|c|}
%\noalign{\hrule height0.8pt}
\hline
 $n$& $m$\\
\hline
 4&3,7,13,17,21,23,39,49,91\\
 6&5,$5^2$,13,41,65,$13^2$,205\\
 8&5,7,11,13,17,25,49,65,77,91,$11^2$\\
 10&9,13,17,81,89,117,$13^2$\\
 12&11,19,23,29,$11^2$,67,209,$19^2$,261\\
14&13,$13^2$,377\\
 16&11,13,17,23,$11^2$,143,187\\
18&17,19,53,137,$17^2$,323,$19^2$\\
20&19,41,$19^2$,779\\
 \hline
%\noalign{\hrule height0.8pt}
\end{tabular}
\end{center}
\end{table}
This table shows that there exist many MDS self-dual code which do
not satisfy the inequality (\ref{eq:boun}).
%\[
%\left(
%\begin{array}{c}{n-1} \\ {n/2-1}\end{array}\right)
%\geq 2^{n/2-1}.

%%%%%%%%%%%%%%%%%%%%  references  %%%%%%%%%%%%%%%%%%%%%%
\section*{Conclusions}
The goal of this work was to survey and provide a unified framework
for codes over chain rings, principal ideal rings, and formal power
series rings. This allowed us to make connections between the
results given in the literature and then extend these results. In
particular, the structure of constacyclic codes over formal power
series and chain rings was given. We also provided necessary and
sufficient conditions on the existence of MDS codes over principal
ideal rings. Further, infinite families of self-dual MDS codes were
are constructed over chain rings, principal ideal rings and formal
power series.
\begin{center}
\textbf{Acknowledgements}
\end{center}
The authors would like to thank Steven Dougherty for his helpful
comments, in particular for motivating the results in Section 4.

\end{document}